\documentclass[aps,prl,twocolumn,superscriptaddress,nofootinbib,nobalancelastpage,showpacs,nobibnotes]{revtex4}
\usepackage{epsfig}

\begin{document}
\newcommand{\identity}{\:\mbox{\sf 1} \hspace{-0.37em} \mbox{\sf 1}\,}

\title
{Evolution speed in some coupled-spin models}

\author{R. F. Sawyer}\email{sawyer@vulcan.physics.ucsb.edu}
\affiliation{Department of Physics, University of California at
Santa Barbara, Santa Barbara, California 93106}

\begin{abstract}

We investigate the time evolution of some models with N spins and pairwise couplings, 
for the case of large N, in order to compare evolution times with ``speed limit" minima 
derived in the literature. Both in a (symmetric) case with couplings of the same strength 
between each pair and in a case of broken symmetry, the times necessary for evolution 
to a state in which the simplest
initial state has evolved into a nearly orthogonal state are proportional to 1/N, as is the speed
limit time. However the coefficient in the broken symmetry case comes much closer to the
speed limit value. Introducing a different criterion for evolution speed, based on macroscopic 
changes in occupation, we find a corresponding enhancement in rates in the asymmetric case as compared
to the symmetric case.

\pacs{03.65.-w, 03.65.Ud, 03.67.-a}

\end{abstract}

\maketitle

\section{Introduction}
In a recent series of papers, Giovannetti, Lloyd, and Maccone \cite{skinny1}-\cite{skinny3} have elucidated
the role of entanglement in approaching  theoretical minima of the
times required for quantum systems to progress from a prescribed initial state to a state that is orthogonal or 
nearly orthogonal to the initial state. The 
criterion for evaluating the outcome is the reduction of the square of the overlap of the initial and final wave-functions 
to some value less than $\epsilon$, where $\epsilon$ is fairly small,
but not necessarily infinitesimal, $P(t) \equiv \langle \Psi(t)|\Psi(0)\rangle <\epsilon$. 
The ``theoretical minimum" of evolution times that we refer to depends on the lesser of the two quantities,
\begin{eqnarray}
\Delta E\equiv\sqrt{ \langle H^2\rangle -\langle H\rangle ^2},
\end{eqnarray}
and $E-E_0\equiv <H>-E_0$, where $E_0$ is the ground state energy of the system and the brackets $<>$ in all cases
stand for the expectation value in the initial state. In terms of these quantities the fastest evolution time for meeting the criterion $P(\tau)=\epsilon$, as derived in refs. \cite{old1}-\cite{old4} and sharpened in refs. \cite{skinny1}- \cite{skinny3}
is given by,\footnote{We take $\hbar=1$ throughout.}

\begin{equation}
\tau \ge\,\,{\rm max}\,\Bigr[{\pi \alpha(\epsilon) \over 2(E-E_0)}, {\pi \beta (\epsilon) \over 2\,\Delta E}\Bigr]\,\,,
\label{speedlimit}
\end{equation}
where,
$\alpha(\epsilon)$ and $\beta(\epsilon)=2{\rm arccos}(\sqrt \epsilon)/\pi $ are functions
that are equal to unity at $\epsilon=0$ and are plotted in ref.\cite{skinny2}.
The time for which the equality holds in (\ref{speedlimit}) will be referred to as the speed limit time. 

As shown in ref. \cite{skinny1}, it is easy to find models in which this 
limit is approached when one chooses initial states in which most of the energy resources are 
concentrated in one subsystem of sufficient simplicity. It also easy to construct coherent combinations
of basis states that are individually of the form of products of subsystem states, 
in order to achieve fast evolution.
But as the authors demonstrate, when the system is composed of multiple simple subsystems with a  
Hamiltonian that is relatively homogeneous across the range of subsystems, and when the system begins in an initial product state, then the interaction must generate entanglement among the subsystems in order to approach the 
speed limit for the evolution. Ref. \cite{skinny2} gives an example in a system of N spins, where one of the
terms in the Hamiltonian is a product of the operators for all spin states,

\begin{equation}
H=\omega_0 \sum_{k=1}^N(1-\sigma_1^{(k)})+\omega (1-\Pi_{k=1}^{N}\sigma_1^{(k)})\, ,
\label{skinny}
\end{equation}
and where the speed limit is realized for choice of parameters such that $\omega>>\sqrt{N} \omega_0$.
In the present work we give some examples of quite different systems that can approach the speed
limit, again involving quite homogeneous Hamiltonians and beginning from
a separable initial state. But our models are based on a more conventional pairwise
coupling among the spins,
 
\begin{equation}
H=G\sum_{i,j}^N \lambda_{i,j}(\sigma_+^{(i)}\sigma_-^{(j)}+\sigma_-^{(i)}\sigma_+^{(j)})\, ,
\label{origham}
\end{equation}
 where $\sigma_+$ and $\sigma_ -$ are the usual raising and lowering operators for an individual spin, and where
we will discuss different choices for the $\lambda_{i,j}$ couplings. 

Within the context of these systems we define an alternative, and rougher, measure of speed of evolution, different from the speed limit rate. This alternative definition might, for some purposes, better characterize the system's rate of undergoing macroscopic changes. 
We illustrate by dividing the spin sites into an ``upper"
tier with $j=1,2..., n_1$ and a ``lower" tier $j=n1+1, n1+2..., n1+n2$, where $n_1+n_2=N$.
We consider initial states
in which all of the upper tier of spins are pointed up and all in the lower tier are
pointed down,
\begin{equation}
|\Psi(0) \rangle =|\uparrow \rangle_1\otimes |\uparrow \rangle_2\otimes ...|\uparrow 
\rangle_{n_1}\otimes  
|\downarrow \rangle_{n_1+1}\otimes ...|\downarrow \rangle_{n_1+n_2}.
\label{initialstate}
\end{equation} 
Then, to define two different kinds of evolution time, we ask the questions:
\begin{enumerate}
\item  
 Following refs \cite{skinny2}, \cite{skinny3}, for a system with the initial state given by (\ref{initialstate}) and governed
 by one of the Hamiltonians that we shall introduce later, what is the time required to evolve to a configuration in which,

\begin{equation}
P(t)=\langle \Psi (t) |Q_1| \Psi (t) \rangle  < \epsilon \,  ,
\label{P1}
\end{equation}

where
\begin{equation}
Q_1=| \Psi(0)\rangle \langle\Psi(0) | \, ,
\end{equation}
and  where $\epsilon$ is a fairly small number? This is the time that defines the speed limit.

\item 
For the same system, what is the time required for the occupancy of, say, the upper tier of states
to evolve from the initial condition of all-spins-up to a condition in which the upper tier has much different
occupancy? To quantify this question, we define
\begin{equation}
R(t)=\langle \Psi (t)| Q_2|  \Psi(t) \rangle
\label{defR}
\end{equation}
where 
\begin{equation}
Q_2=(2 n_1)^{-1}\sum_{i=1}^{n_1}(1+\sigma _3^{(i)}),
\end{equation} and ask, for
example, for the time at which $R(t)=.5$. The function $R$ is equal to the average fraction of the upper-tier spins
that remain pointed up.

\end{enumerate}
For the Hamiltonian of (\ref{skinny}) in the limit of parameters for which the speed limit is approached, the two times defined by the above questions are of the same
order of magnitude. But for our Hamiltonians, they can differ greatly, one from the other, in a way that appears to depend on the symmetry of the Hamiltonian. 

Before turning to specific cases we note that in the first 
definition of evolution time it would suffice for a single spin in the system to be flipped to obtain an 
orthogonal configuration. For our case of a system of a large number of spins this would change the average 
occupancy of the upper tier almost not at all. Alternatively, we could think of evolution
to a product wave-function in which each upper-tier single-spin state, originally spin-up, is $4$ \% mixed (in probability) 
with the spin-down state, so that if we had $100$ states in the upper tier the overlap, $P$, that enters the
first criterion, above, would be $\approx \exp(-4)$; whereas, looking at the second criterion, 
the mixing of the entire upper tier would be only $4$\%. In this case the speed-limit time is much shorter
than the mixing time. 

We shall  be particulary interested in the limit of a large number, $N$, of spins, in our pair-interaction models. The speed
limit time $\tau$ will turn out to be proportional to $N^{-1}$ in these systems, keeping coupling constants fixed.

\section{Models and their speed limits}

In what follows we will consider two possible choices for the coupling-constant function in
(\ref{origham}), $\lambda_{i,j}$. 

Case A: $\lambda_{i,j}=1/2$. This case has complete permutational symmetry among the spins.
We can write the Hamiltonian as,
\begin{equation}
H_A= {1 \over 2} G (J_+ J_- +J_-J_+) \, ,
\end{equation}
 where
\begin{equation}
J_\pm=\sum_i \sigma^{(i)}_\pm \, ,
\end{equation}
are the raising and lowering operators for the total spin of the system.
We specialize
to the case $n_1=n_2 \equiv n$,
for simplicity. Then we obtain the values: $E_{0}=0$, $E=n G$, $\Delta E=\sqrt 2 nG$ . 
In this case the speed limit will be thus given by the term involving $E-E_0$ in (\ref{speedlimit}),

\begin{equation}
\tau_A \ge \,{\pi \alpha(\epsilon) \over 2 n G}\,\,.
\label{sla}
\end{equation}

Case B:   We let the upper set of spins interact with the lower set exactly as in case A, but take no interactions between
pairs of spins both in the upper tier, or both in the lower tier. Taking $\lambda_{i,j}=1/2$ for all upper-lower
connections and  $\lambda_{i,j}=0$ for the intra-tier connections, and defining,
\begin{eqnarray}
K_\pm=\sum_{j=1}^{n} \sigma^{(i)}_\pm \, ,
\nonumber\\
L_\pm=\sum_{j=n+1}^{2n} \sigma^{(i)}_\pm \, ,
\end{eqnarray}
the Hamiltonian of (\ref{origham}) is,

\begin{equation}
H_B=G (K_+ L_- +K_-L_+) .
\label{hamB}
\end{equation}
In this case we readily find $\Delta E= nG$ and E=0. Determination of the ground state energy $E_0$ requires 
a numerical calculation which we discuss later. It will turn out that in this case the speed limit is determined
by the $\Delta E$ term in (\ref{speedlimit}),
\begin{equation}
\tau_B \ge \,{\pi \beta(\epsilon) \over 2 n G}\,\, .
\label{slb}
\end{equation}

Before solving  these models we consider some possible outcomes for the case of large $n$. 
Since $G$ is the only dimensional quantity in the theory, characteristic times are of the form,

\begin{equation}
\tau=[G\, h(n)]^{-1} \, . 
\end{equation}

For the evolution to proceed at a finite fraction of the speed limits (\ref{sla}) and (\ref{slb}) requires $h(n)={\rm const}\times n$ for either case. 
For small times the simple perturbative probability for a single spin to have changed its state is of the form, 
\begin{equation}
{\rm prob}={\rm const.}\times G^2 t^2 n\,\, ,
\label{pert}
\end{equation}
 giving a characteristic time for appreciable change of single-spin occupation probability of,
\begin{equation}
 \tau_2={\rm const.} \times (G \, \sqrt{n})^{-1} \, .
\label{pert2}
\end{equation}
In this kind of model, as we indicated earlier, reaching a given $P(t)<\epsilon$ in the case of large $n$
requires only a change of single state probability of order ${\rm const.} /n$, and the perturbative time to realize this
change is, from (\ref{pert}) of order $\tau_1={\rm const.}\times  (Gn)^{-1}$. Thus we expect to get the
same dependence on $G$ and $n$ as in the above speed limit times, but with a constant that must be calculated.
In contrast the perturbative estimate of the characteristic time for change in $R(t)$ is $\tau_2=
{\rm const.} \times (G \, \sqrt{n})^{-1}$. When we find, for example, that the evolution for large $n$ is faster,
for example going like $1/n$ or $\log(n)/n$, we shall refer to the phenomenon as a ``speed-up".

\section{Solution for model A}

We take $n_1=n_2= n$ to define the initial
configuration, and we rewrite $H_A$ with $\vec J=\vec K+\vec L$ as,
\begin{equation}
H_A=G[(\vec K +\vec L)\cdot (\vec K +\vec L)- (K_3+L_3)^2] \, .
\label{newHA}
\end{equation}
Since we are beginning in an eigenstate of $(K_3+L_3)$, which is
conserved and equal to zero, we can discard the final term on the RHS of
(\ref{newHA}). Next we note that $\vec K \cdot \vec K$ and $\vec L \cdot \vec L$
are separately conserved under $H_A$. The initial state is an eigenstate of these
two operators with eigenvalues $(n/2)(n/2+1)$. The eigenvalues of $H_A$ are
given by $G j(j+1)$, associated with the states in which the angular momentum
$\vec K$ has been added to the angular momentum $\vec L$ to give the total
angular momentum quantum number, $j$. Thus to follow the time evolution
it is only necessary to resolve the initial state into eigenstates of $\vec J\cdot \vec J$,
and let the system develop for some period of time. We then take the expectation value
of whatever operator, $Q$, we wish to represent the mixing of the system. 
To address the
criterion \#1 (i.e. to compare with the speed limit) we calculate $P(t)$, defined in (\ref{P1}),
\begin{equation}
P(t)=\Bigr \vert \sum_{j=0}^{n} e^{-iG\,j(j+1)t}[C_{n/2,\,\,n/2,\,\,j}^{-n/2,\,n/2,\,0}]^2 \Bigr \vert ^2 \, ,
\label{P(t)} 
\end{equation}
where we have used the notation, $C_{j_1,j_2,j}^{m_1,m_2,m}$ for the Clebsch-Gordan coefficient.

For large $n$ and for $t<b G ^{-1} n^{-1/2}$, where $b$ is some number smaller than unity,
 a sufficiently accurate asymptotic estimate of  the Clebsch-Gordan coefficients to be used in (\ref{P(t)}) 
 is given by,
\begin{equation}
\Bigr [C_{n/2,\,\,n/2,\,\,j}^{-n/2,\,n/2,\,0}\Bigr]^2 \approx \Bigr ( {j \over n }\Bigl ) 
\exp[-{j^2 \over 2n}] \, ,
\end{equation}
leading directly to the result,
\begin{equation}
P(t)=(1+G^2 n^2 t^2)^{-1}(1+O(G \sqrt n \, t) \, ,
\label{caseAspeed}
\end{equation}
which holds for times less than those of order $(G \sqrt n)^{-1}$. We have confirmed the limiting
result (\ref{caseAspeed}) with direct numerical calculations. In fig.1 we plot the function $P$
against the dimensionless coordinate, $n\,G\,t$. On the same plot we indicate the speed limit rate implied by (\ref{sla}),
plotting the function $\epsilon (t)$, where $t=\pi \alpha [\epsilon(t)]/ 2(E-E_0)$, $E-E_0=nG$, and we have used the
approximation $\alpha(\epsilon) \approx \beta(\epsilon) ^2$ suggested in ref.\cite{skinny3}. We see 
from fig.1 that the evolution in the model procedes at roughly half the speed-limit rate. 

\begin {figure}[ht]
    \begin{center}
        \epsfxsize 2.75in
        \begin{tabular}{rc}
            \vbox{\hbox{
$\displaystyle{ \, { } }$
               \hskip -0.1in \null} 
} &
            \epsfbox{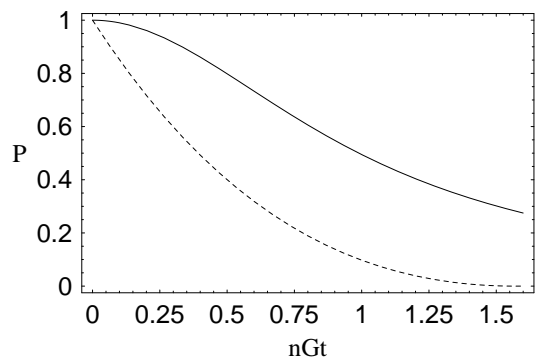} \\
            &
            \hbox{} \\
        \end{tabular}
    \end{center}
\label{fig.1}
\protect\caption
    {%
The time evolution of the function $P(t)$ for model A (solid line), plotted as a function of the 
dimensionless variable, $Gnt$. The dashed curve shows the evolution that saturates the 
speed limit.
 }
\end {figure}

We have confirmed the analytic approximations leading
to (\ref{caseAspeed}) by direct numerical calculations, and find that the curves for $P$ are essentially
congruent for all $n\ge 4$.
A numerical calculation over a longer time interval shows that after
a time of order $G^{-1} n^{-1/2}$, where the approximations of (\ref{caseAspeed}) fail, the function
$P(t)$ remains at a small value for an extended time that is of order $(Gn)^{-1}$, then reversing its earlier
decline in a symmetric fashion; the function is periodic with a period $2 \pi /G$, as is clear from (\ref{P(t)}).

To address criterion \#2, based on the evolution of the expectation value of the
z-component of the total spin of the upper tier, we calculate $R(t)$ defined in (\ref{defR}), obtaining,

\begin{eqnarray}
& R(t)=1-\sum_{k=0}^{n} {k \over n} \Bigr [
\nonumber\\
 \Bigr \vert \, &\sum_{j=0}^{n} e^{-iG\,j(j+1)t}C_{n/2,\,\,n/2,\,\,j}^{-n/2+k,\,n/2-k,\,0}
C_{n/2,\,\,n/2,\,\,j}^{-n/2,\,\,n/2,\,\,0}\, \Bigr \vert \,^2 \Bigr ] \, .
\nonumber\\
\,
\label{symmetriccase}
\end{eqnarray}

In fig.2, we show plots of numerical results for $n=8,16,32,64,128,256$, in the region
in which $R(t)$ declines from unity to a value less than $.6$. 

\begin {figure}[ht]
    \begin{center}
        \epsfxsize 2.75in
        \begin{tabular}{rc}
            \vbox{\hbox{
$\displaystyle{ \, { } }$
               \hskip -0.1in \null} 
} &
            \epsfbox{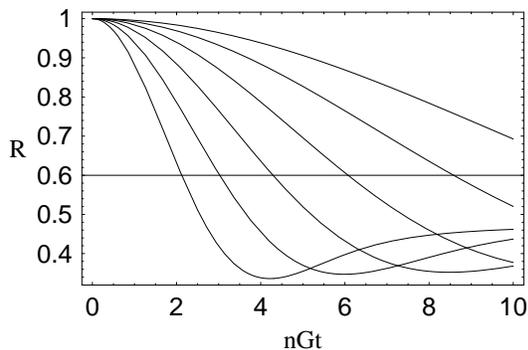} \\
            &
            \hbox{} \\
        \end{tabular}
    \end{center}
\label{fig.2}
\protect\caption
    {%
The time evolution of the function $R(t)$ for model A, plotted as a function of the 
dimensionless variable, $Gnt$, for values $n=8,16,32,64,128,256$. The curves for
higher values of $n$ lie progressively to the right of the lower $n$ curves. 
 }
\end {figure}

The time required to reach $R(Gnt)=.6$ is strongly dependent 
on $n$, as we see in fig.2. If $R$ were to have developed at a fixed fraction of the speed limit rate
(as did $P$ in the previous example), then the value of the dimensionless
variable $Gnt$ required to reach this point would have been independent of $n$ . Friedland and Lunardini \cite{fl2} have established that for large $n$ the evolution time 
behaves as $(G \sqrt n)^{-1}$, with a coefficient
of order unity; so that the characteristic values of  $Gnt$ \underline{increase} as $\sqrt{n}$ .  This is essentially the perturbative time dependence that we defined above \footnote{This result is again obtainable by using the asymptotic 
forms of Clebsch-Gordan coefficients. In place of $\langle|Q|\rangle$, Friedland and Lunardini consider the probability that a single,
typical spin is flipped, summed over all possible fates for the remaining system, and ingeniously reduce the problem
in such a way that only a single sum analogous to that in (\ref{symmetriccase}) need be performed.}, and is already
roughly borne out in a fit to the intercepts of the curves shown in fig. 2 
with the line $R=.6$.

\section{Solutions for model B}
First we enumerate the set of states that are dynamically connected, given the initial state (\ref{initialstate}) with $n_1=n_2=n$ and the Hamiltonian
(\ref{hamB}). The initial state is an eigenstate of $\vec K^2 $, and $\vec L^2$, each of which commutes with the Hamiltonian, and each with the same eigenvalue, $(n/2)(n/2+1)$. The total $\hat z$ component ``angular momentum", $K_3+L_3$ is 
also conserved and equal to zero for our set of states.
Any number of the $n$ spins in the upper tier, all initially up, may be flipped, leading to $ n+1$ possibilities for
$K_3$ for this tier by itself. For each such configuration there is also only a single state of the lower tier. Thus we can index the states by the number of flips plus one, $i$, where $i$ takes on the values $1,2...n+1$. We express the operator products that occur in the Hamiltonian in this basis as,

\begin{eqnarray}
\langle i | K_-L_+| i-1\rangle = (n-i+1)(i) ;\,\,i=1,2...n+1
\nonumber\\
\langle i-1 |K_+L_-| i\rangle=(n-i+2)(i-1);\,\,i=1,2...n+1\,\, ,
\nonumber\\
\,
\label{bmats}
\end{eqnarray}
which come directly from the standard angular momentum matrices. We now solve numerically for 
a $(n+1)$ component wave function $\Psi(t)$, using the Hamiltonian (\ref{hamB}) with the substitution (\ref{bmats})
and the initial condition 
$\Psi_i(0)=\delta_{i,1}$. We then calculate both of our measures of change,
\begin{equation}
P (t)=|\Psi_1 (t)|^2\,\, ,
\end{equation}
and
\begin{equation}
R(t)=\sum_{i=1}^{n+1} |\Psi_i(t)|^2 (n-i+1)n^{-1}\,\, .
\end{equation}

We also solve numerically for the ground state energy, to confirm that it is the $\Delta E$ condition of(\ref{speedlimit})
that applies. We find that the approximation,
\begin{equation}
E_0=-0.34 n - 0.49 n^2\,\, ,
\end{equation}
fits the energies to within 1\% over the range $n=4-100$. Thus we have $E-E_0>\Delta E$ over the
whole range that we consider, and the speed limit is determined from the term in (\ref{speedlimit}) 
involving $\Delta E$.
In fig.3 we plot $P(t)$ for $n=64$ against the variable $gNt$. On the same plot we indicate the limitation imposed by (\ref{slb})
by plotting the function $\epsilon (t)$ that is the solution to the equation $t=\pi \beta (\epsilon )/ (2\, \Delta E)$
where $\Delta E=nG$. Comparing with the case of model A, we see that the evolution proceeds at much more 
nearly the speed limit rate. The plots of $P$ are almost identical for all $n \ge 4$.

\begin {figure}[ht]
    \begin{center}
        \epsfxsize 2.75in
        \begin{tabular}{rc}
            \vbox{\hbox{
$\displaystyle{ \, { } }$
               \hskip -0.1in \null} 
} &
            \epsfbox{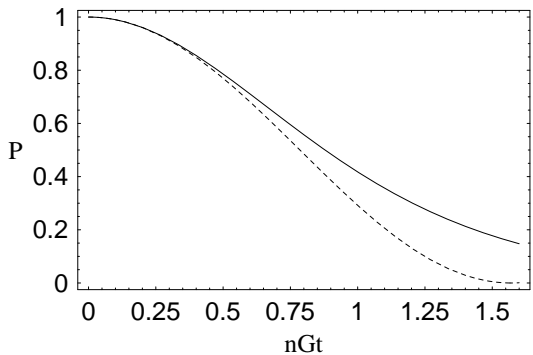} \\
            &
            \hbox{} \\
        \end{tabular}
    \end{center}
\label{fig.3}
\protect\caption
    {%
The time evolution of the function $P(t)$ for model B (solid line), plotted as a function of the 
dimensionless variable, $Gnt$. The dashed curve shows the evolution that saturates the 
speed limit.
 }
\end {figure}

In fig 4. we show  plots of the scaled function $R(t)$ for $n=8,16,32,64,128,256,512$, again plotting against
the variable $Gnt$. 
Note that the intersections with the line $R=.4$ occur at very nearly equally spaced points. Equal
spacing indicates a mixing time decreasing with $n$ as $\tau \propto {\rm log} (n)/n$. This is in 
contrast to the symmetrical
model A in which characteristic time decreases as $\tau \propto 1/\sqrt n$, which we defined as the
``perturbative" time. Thus we have
a large speed-up over the perturbative estimate, as $n$ increases. For the case $n=128$, if we define the 
characteristic time according to criterion \#2, the speed-up over the perturbative rate is a factor of approximately 4. 

\begin {figure}[ht]
    \begin{center}
        \epsfxsize 2.75in
        \begin{tabular}{rc}
            \vbox{\hbox{
$\displaystyle{ \, { } }$
               \hskip -0.1in \null} 
} &
            \epsfbox{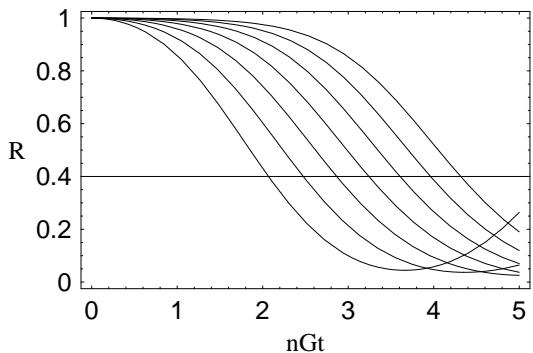} \\
            &
            \hbox{} \\
        \end{tabular}
    \end{center}
\label{fig.4}
\protect\caption
    {%
The time evolution of the function $R(t)$ for model B, plotted as a function of the 
dimensionless variable, $Gnt$, for values $n=8,16,32,64,128,256,512$. The curves for
higher values of $n$ lie progressively to the right of the lower $n$ curves. 
 }
\end {figure}

\section{Discussion}We have considered large spin systems with couplings among pairs such that  the quantum speed limit
time is proportional to the inverse of the number of spins, $\tau \propto N^{-1}$, where we taken an initial state which
is completely separable. We have shown that the time required for the squared overlap , $P(t)$, of the evolved state with the initial state
to become small
is indeed proportional to $N^{-1}$, with a coefficient which in model A is about twice the speed limit value, and in model B
is only 10-20\% greater than the speed limit time, depending on the value of $\epsilon$ one chooses for purposes of comparison.

We have also looked at the time scale for change of another attribute of these systems, $R(t)$, the fraction of upper tier spins
that have been flipped. Here the
results were drastically different for the two models; in model A, where there were couplings among all of the spins,
the evolution time for $R(t)$ is proportional to $N^{-1/2}$, which we characterize as the ``perturbative" time, rather than to $N^{-1}$. But in model B, with breaking of permutational symmetry\footnote{Note that we broke the spin states into
two sets both for the purpose of breaking the permutation symmetry by setting to zero the intraset interactions, and for the purpose
defining which states were initially up and which down. We used the same division for the two purposes only to be able
(for large $N$) to reduce the evolution problem to calculable dimensions. This is an example of how very minimal is our breaking of the 
symmetries of the symmetric
model.} the evolution time is reduced to
a time of order $N^{-1} {\rm log}[N]$, a speed-up to near the $N^{-1}$ level of the speed-limit time.
It is tempting to conjecture that with more breaking of the symmetry we could lose the log, and
perhaps even approach speed-limit rates. The curves shown in ref. \cite{brs}, however, for the case
of completely broken symmetry, were inconclusive in this regard, due to computational limitations. More generally, it would seem to us to be of some interest first to define ``macroscopic change"
more precisely in a large N limit, and then to try to find general results, analogous to the speed-limit
results, on the limitations for the rates for this macroscopic change. 

We should note that we have tested models in which $n_1 \ne n_2$ and find results qualitatively the same as those reported for the two models treated in this paper. We set out, in calculating $R(t)$ to get an estimate of the rate at which the average total $\hat z$ component spin of the upper tier, $K_3$, 
changes. But see that in model B the measure $R(t)$ is very nearly zero at the first minimum, indicating that
almost \underline{all} the spins have been flipped. We could compare the results, where (up to logarithms) the characteristic time for evolution is 
of order $(G N)^{-1}$,
to the results for the simplest Hamiltonian that will flip each spin in time $(G N)^{-1}$, namely,
\begin{equation}
H_C={\pi \over 2}G N \sum_{i=1}^N \sigma_1^{(i)}\,\, .
\end{equation}
Measuring the ``energy resources" of this Hamiltonian by $\Delta E=
(\pi /2) G N^{3/2}$, and comparing to $\Delta E= G(N/2)$ for the Hamiltonian (\ref{hamB}), we conclude
that the model with pair couplings and consequent many-body entanglements does indeed succeed
in rotating the complete ensemble of spins in a very efficient way, for large $n$.

Finally, we comment on the question as to how much substance is contained in the statement that
``entanglement is the key to speed-up of rates" a paraphrase of a point emphasized in \cite{skinny1} and also in
a quite different context in \cite{brs}. Of course, states do become entangled during the evolution under the 
interactions that we have taken in our models. But there is no general agreement known to us on the 
quantification of entanglement in the case of multicomponent systems, even in the case of pure states. We can, however,
perhaps rather arbitrarily, consider the entanglement of two systems, $U$ and $L$, where $U$ contains all of the upper tier spins and 
$L$ contains the lower tier states. Then a measure of entanglement between the two systems is given by the \it entropy of
entanglement \rm obtained by tracing out one subsystem's coordinates in the density matrix, say the $U$ subsystem, to define a reduced 
density matrix $\rho_L$ for the $L$ subsystem, and then calculating the von-Neumann entropy corresponding to $\rho_L$ \cite{bennett} \cite{wootters},

\begin{equation}
S_e=-Tr \rho_L\,\log[\rho_L]\,\, .
\end{equation}

We have calculated $S_e$ as a function of time for models A and B, using the wave-function solutions
underlying the calculations discussed above. At a time $Gnt=1$, at which $P\approx .5$ in either model,
we find that, so defined, the entanglement is about 30\%
more for the case with broken symmetry than for the completely symmetric case, and relatively independent of
the number of particles. Thus in our 
models there is an perhaps an indication of a correlation between the rate of growth of entanglement
and the efficiency of transformation. Of course, in the pure states that we have considered, the entanglement, as defined above, can
both increase and decrease in time, and comparisons will depend on the time of the sampling.
\section{Appendix}


\begin{thebibliography}{99}
\bibitem{skinny1}V. Giovannetti, S.Lloyd, L. Maccone, Europhys. Lett. 62, 615 (2003), quant-ph/0206001
\bibitem{skinny2}  V. Giovannetti, S.Lloyd, L. Maccone, Phys. Rev. {\bf A67}, 052109 (2003), quant-ph/0210197
\bibitem{skinny3} V. Giovannetti, S.Lloyd, L. Maccone, Fluctuations and Noise in Photonics and Quantum Optics. 
Edited by Abbott, D. et al. Proceedings of the SPIE, Volume 5111, pp. 1-6 (2003), quant-ph/0303085
\bibitem{old1}K. Bhattacharya, J. Phys. {\bf A16}, 2993 (1983)
\bibitem{old2}P. Pfeifer, Phys. Rev. Lett. {\bf 70}, 3365 (1993)
\bibitem{old3}L. Mandelstam and I. G. Tamm, J. Phys. USSR {\bf 9}, 249 (1945)
\bibitem{old4}N. Margulus and L. B. Levitan, Physica {\bf D120}, 188 (1998)
\bibitem{brs} N. F. Bell,  A. A. Rawlinson, and R. F. Sawyer, Phys.Lett. {\bf B573}, 86 (2003) 
\bibitem{fl2}A. Friedland and C. Lunardini,  JHEP{\bf 0310}, 043 (2003) ;quant-ph/0303085
\bibitem{bennett} C. H. Bennett, H. J. Bernstein, S. Popescu and B. Schumacher,
Phys.\ Rev.\ A {\bf 53}, 2046 (1996).
\bibitem{wootters} S. Hill and W. K. Wootters, Phys.\ Rev.\ Lett.\ {\bf 78}, 5022
(1997); W. K. Wootters, {\it ibid.} {\bf 80}, 2245 (1998)
\end{thebibliography}
\end{document}